\documentclass[
aps,
prb,
reprint,
showkeys,
twocolumn,
nofootinbib,
superscriptaddress,
]{revtex4-2}

\usepackage{natbib}
\usepackage{tabularx}
\usepackage{textgreek}
\usepackage{subscript}
\usepackage{natbib}
\usepackage{amssymb}
\usepackage{color}
\usepackage{amsfonts}
\usepackage{textcomp}
\usepackage{amsmath}
\usepackage[charter]{mathdesign}
\usepackage{easyReview}
\usepackage{microtype} 
\usepackage{upgreek}
\usepackage{subfigure}
\usepackage{booktabs}
\newcommand{\ped}[1]{\ensuremath{_{\rm #1}}}
\newcommand{\apex}[1]{\ensuremath{^{\rm #1}}}
\definecolor{link}{RGB}{57,106,177}
\definecolor{darkgreen}{RGB}{0,128,0}
\definecolor{blue}{RGB}{0,0,0}
\usepackage{float}
\usepackage{graphicx}
\usepackage{hyperref}
\hypersetup{
	colorlinks=true,
	linkcolor=link,
	citecolor=link,
	urlcolor=link
}

\begin{document}

\title{Facile synthesis of palladium hydride via ionic gate-driven protonation using a deep eutectic solvent}
	
\author{Gaia Gavello}
\affiliation{Department of Applied Science and Technology, Politecnico di Torino, 10129 Torino, Italy}
\author{Giorgio Tofani}
\affiliation{Department of Pharmacy, University of Pisa, 56126 Pisa, Italy}
\affiliation{Department of Physics, University of Pisa, 56126 Pisa, Italy}
\affiliation{Department of Catalysis and Chemical Reaction Engineering, National Institute of Chemistry, SI-1001 Ljubljana, Slovenia}
\author{Domenico De Fazio}
\affiliation{Department of Molecular Sciences and Nanosystems, Ca' Foscari University of Venice, 30172 Venice, Italy}
\affiliation{ICFO-Institut de Ciencies Fotoniques, The Barcelona Institute of Science and Technology, 08860 Castelldefels (Barcelona), Spain}
\author{Stefania Lettieri}
\affiliation{Department of Applied Science and Technology, Politecnico di Torino, 10129 Torino, Italy}
\author{Andrea Mezzetta}
\author{Lorenzo Guazzelli}
\affiliation{Department of Pharmacy, University of Pisa, 56126 Pisa, Italy}
\author{Christian S. Pomelli}
\affiliation{Department of Pharmacy, University of Pisa, 56126 Pisa, Italy}
\author{Renato S. Gonnelli}
\author{Erik Piatti}
\email{erik.piatti@polito.it}
\author{Dario Daghero}
\affiliation{Department of Applied Science and Technology, Politecnico di Torino, 10129 Torino, Italy}


\begin{abstract}
\noindent Developing novel protocols for hydrogen (H) loading is crucial for furthering the investigation of hydrides as potential high-temperature superconductors at lower pressures compared to recent discoveries.
Ionic gating-induced protonation (IGP) has emerged as a promising technique for H loading due to its inherent simplicity, but it can be limited in the maximum density of injected H when ionic liquids are used as a gating medium.
Additionally, most ionic liquids are limited by high production costs, complex synthesis, and potential toxicity.
Here, we demonstrate that large H concentrations can be successfully injected in both palladium (Pd) bulk foils and thin films (up to a stoichiometry PdH\ped{0.89}) by using a deep eutectic solvent (DES) choline chloride:glycerol 1:3 as gate electrolyte and applying gate voltages in excess of the cathodic stability limit.
The attained H concentrations are large enough to induce superconductivity in Pd, albeit with an incomplete resistive transition which suggests a strongly inhomogeneous H incorporation in the Pd matrix.
This DES-based IGP protocol can be used as a guideline for maximizing H loading in different materials, although specific details of the applied voltage profile might require adjustments based on the material under investigation.
\end{abstract}


\maketitle
\noindent
\section{Introduction}
Recent breakthroughs in achieving high-temperature superconductivity (HTSC) in high-pressure hydrides ($\gtrsim100$\,GPa) have reignited interest in exploring hydrogen (H)-rich materials\,\cite{Drozdov2015Nat,Drozdov2019Nat}.
These discoveries stimulate the search for novel hydrides where H-induced, high-frequency vibrational modes could trigger HTSC at considerably lower pressures than those currently required.
Identifying such compounds would represent a significant advancement toward more feasible and environmentally sustainable superconducting technologies\,\cite{Boeri2022IOP}.
In this pursuit, establishing efficient and reliable techniques for H incorporation into promising candidate materials is crucial.\\
\indent Conventional techniques for attaining H charging include gas absorption from high-pressure gas\,\cite{Syed2016PSSC, Kawae2020JPSJ, Hemmes1989RSI, SCHIRBER_1973, Kato_2024}, electrochemical loading\,\cite{Syed2016PSSC, Kawae2020JPSJ, Skoskiewicz1972, HARPER_1974, Krahn_1978}, and ion implantation\,\cite{Stritzker1972,OCHMANN_1983, Syed2016PSSC, Kawae2020JPSJ, Heim_1975}.
Among electrochemical loading techniques, ionic gating-induced protonation (IGP)\,\cite{Ji_2012, Shimizu_2013, Lu2017, Wang_2017, Cui_2019, Rafique2019Nano, Tan_2019, Li_2019ADVSCI, Ri_2020, Li_2020, Shen_2021, Meng2022PRB, Liu_2022, Piatti2023CommunPhys, Wang_2019, Guo_2022, Chu_2022, Prando2023PRM, Wu2023, Wu_2024ACS, Yan_2024} has emerged as a particularly attractive approach due to its inherent simplicity.
Notably, this method involves applying a voltage difference across an electrolyte in direct contact with the target material, as sketched in Fig.\,\ref{fig:1}a.
The resulting strong electric field ($\gtrsim1$\,MV\,cm\apex{-1}) established at the sample-electrolyte interface triggers electrochemical processes that generate H\apex{+} ions, which are subsequently driven into the material\,\cite{Lu2017, Boeri2022IOP, Piatti2023CommunPhys, Meng2022PRB}.
Moreover, the entire IGP apparatus can be readily integrated into a cryostat, enabling rapid cooling to prevent H desorption and allowing simultaneous \textit{in-situ} characterizations.\\
\indent Previous studies have demonstrated the effectiveness of IGP in modifying the electronic properties of materials by H injection\,\cite{Ji_2012, Shimizu_2013, Lu2017, Wang_2017, Cui_2019, Rafique2019Nano, Tan_2019, Li_2019ADVSCI, Ri_2020, Li_2020, Shen_2021, Meng2022PRB, Liu_2022, Piatti2023CommunPhys, Wang_2019, Guo_2022, Chu_2022, Prando2023PRM, Wu2023, Wu_2024ACS, Yan_2024}, including inducing superconductivity in previously non-superconducting materials\,\cite{Cui_2019, Piatti2023CommunPhys, Liu_2022} \add{and attaining an efficient control of magnetism in both platinum\,\cite{Shimizu_2013, Chu_2022} and cobalt\,\cite{Tan_2019} thin films}.
The use of ionic liquids (ILs) as electrolytes enables the traces of water present in them to be used to generate O\apex{2-} and H\apex{+} ions\,\cite{Lu2017}.
Due to their low melting temperature, commercial availability, large electrochemical window and low water uptake, ILs with fluorinated TFSI and BF\ped{4} anions were the most widely used\,\cite{Ji_2012, Shimizu_2013, Lu2017, Wang_2017, Cui_2019, Rafique2019Nano, Tan_2019, Li_2019ADVSCI, Ri_2020, Li_2020, Shen_2021, Meng2022PRB, Liu_2022, Piatti2023CommunPhys, Wang_2019, Guo_2022, Chu_2022, Prando2023PRM, Wu2023, Wu_2024ACS, Yan_2024}.
However, ILs with BF\ped{4} anion hydrolyze to form highly corrosive hydrofluoric acid in the presence of water\,\cite{Freire2010JPCA}, whereas ILs with TFSI anion are very expensive, environmentally unfriendly and highly toxic\,\cite{Singh_2020, Amde_2015, De_2022}.
Additionally, IL-based IGP can lead to the intercalation of the organic ions comprising the IL instead of the H ions\,\cite{Zhang2020NatPhys, Rousuli20202DMater, Wang2021CPB, Piatti2022Nanomat}, and the dependence of IL-based IGP on water electrolysis for H generation limits its efficacy in non-humid environments.\\
\indent Nowadays, deep eutectic solvents (DESs) offer an attractive alternative to ILs due to their lower toxicity, their biodegradability, and the straightforward synthesis process from readily available and inexpensive components\,\cite{Plotka_2020, Hansen_2020, Radovsevic_2015, Khandelwal_2016, Afonso2023GC}.
DESs are mixtures of two or more compounds that exhibit a negative deviation from the ideal eutectic in terms of melting temperature.
It is therefore possible to design a specific DES for a specific application by changing the chemical nature of the compounds and their molar ratio. 
To the best of our knowledge, DESs have never been used in IGP, and their use could make it possible to solve the main critical issues of the usage of ILs.
Additionally, unlike ILs, DESs could generate H through direct solvent dissociation, thereby reducing their dependence on specific environmental conditions.
However, a critical prerequisite for their application in IGP is the demonstration of their ability to promote large H injection.\\
\indent To assess the potential of DESs for IGP, this study investigates H loading into palladium (Pd) using a choline chloride:glycerol 1:3 DES.
This DES features good ionic conductivity due to the presence of the choline chloride salt, and simultaneously acts as a good source of H thanks to glycerol, a species which can dissociate and release H\apex{+} at relatively low potentials\,\cite{Lee2016GC, Lin2017IJHE, Wang2023choline, Li_2016}.
\add{Furthermore, compared to other hydrogen bond donors currently used in the preparation of DES, glycerol is obtained directly from natural sources.}
\add{The molar ratio of choline to glycerol was chosen to minimize the viscosity of the medium. At a molar ratio of 1:3, the DES exhibits a lower viscosity than 1:2 and 1:4, giving potential advantages in terms of conductivity and homogeneity of the H intake\,\cite{Mero2023}.}\\
\indent\add{Among metal elements,} Pd was selected as a model system due to the extensive characterization of the PdH$_x$ hydride and its well-established superconducting properties\,\cite{Skoskiewicz1972, Ganguly_1973, Bennemann1973, Miller_1974, Mitacek_1963, Araki_2004, Standley1979SSC}.
\add{With respect to other metal hydrides,} the direct correlation between absorbed H content and PdH$_x$ electrical resistivity\,\cite{Zhang2002JEC, Flangan_Lewis_1961, Smith_Otterson_1970, Barton_Lewis_1962, Baranowski_1983, Ho_1968} enables the straightforward monitoring of the IGP process.
Additionally, the occurrence of a superconducting transition at ambient pressure for high H concentrations ($0.8 \lesssim x \leq 1$) with a critical temperature up to 9\,K depending on the concentration\,\cite{Skoskiewicz1972, Skośkiewicz_1973, SCHIRBER_1973, Stritzker1972, HARPER_1974, OCHMANN_1983, Krahn_1978, Araki_2004, Standley1979SSC}, provides a clear indicator of large H incorporation into the Pd lattice,
and the well-known tendency of Pd to readily release absorbed H allows for the assessment of the robustness of IGP against potential desorption losses\,\cite{Stern_1984}.
\add{
Another reason to choose Pd is that Pd-noble metal alloys are known to become superconducting when loaded with H\,\cite{Stritzker1974} and a high-temperature superconducting phase has been predicted to occur in PdCuH$_x$\,\cite{Vocaturo2022}. Therefore, the same method used here for pure Pd could be readily applied to these materials in the search for high-temperature H-based superconductors at ambient pressure.
}

\section{Experimental methods}
\subsection{Fabrication of the palladium samples}\label{methods:samples}
Bulk Pd foils were commercially purchased from Sigma Aldrich (99.9\% purity, \add{25\,$\upmu$m nominal thickness}) and used as-is after 15\,min sonication in ethanol.
Thin Pd films were grown on sapphire substrates by photolithography, Pd evaporation and lift-off.
Lithography was performed by using a Heidelberg maskless aligner (MLA150) system. Devices were patterned in the shapes of Hall bars on an AZ ECI 3007 photoresist mask spin-coated at 4000\,rpm (thickness $\approx 0.7\,\upmu$m) and baked at 110\,$^\circ$C for 90\,s.
The development in AZ 726 MIF at room temperature lasted 60\,s.
Ti (nominal 2\,nm) and Pd (nominal 35\,nm) were then sequentially deposited with a Leybold Univex 350 thermal evaporator.
The evaporation was followed by an overnight lift-off in acetone, dipping in isopropanol (2 minutes) and nitrogen blow-drying.
\subsection{AFM characterization}\label{methods:AFM}
Atomic force microscopy measurements were carried out in ambient conditions using a Bruker Innova microscope in tapping mode, with commercial silicon tips having a nominal resonance frequency of 300\,kHz (RTESPA-300).
Image analysis and post-processing were performed using Gwyddion software\,\cite{Gwyddion}.
\subsection{Synthesis of the deep eutectic solvent}\label{methods:DES}
Choline chloride ($>98\%$) and glycerol ($>99\%$) were purchased in 100\,g from Sigma Aldrich.
The DES was prepared using a previously reported procedure\,\cite{Mero2024GC}.
Before preparation, choline chloride and glycerol were dried under vacuum for 6\,h at 80\,$^\circ$C. 
The choline chloride (1 equiv.) and glycerol (3 equiv.) were mixed, and the mixture was stirred at room temperature until a homogeneous transparent liquid was formed.
After the formation, no purification step was needed, and the solvents were kept at room temperature in sealed vessels until their use.
\add{The water content was measured using a Karl Fischer apparatus and resulted in 520\,ppm.}
\subsection{Ionic gating-induced protonation}\label{methods:IGP}
\begin{figure*}[t]
    \centering\includegraphics[width=\textwidth, keepaspectratio]{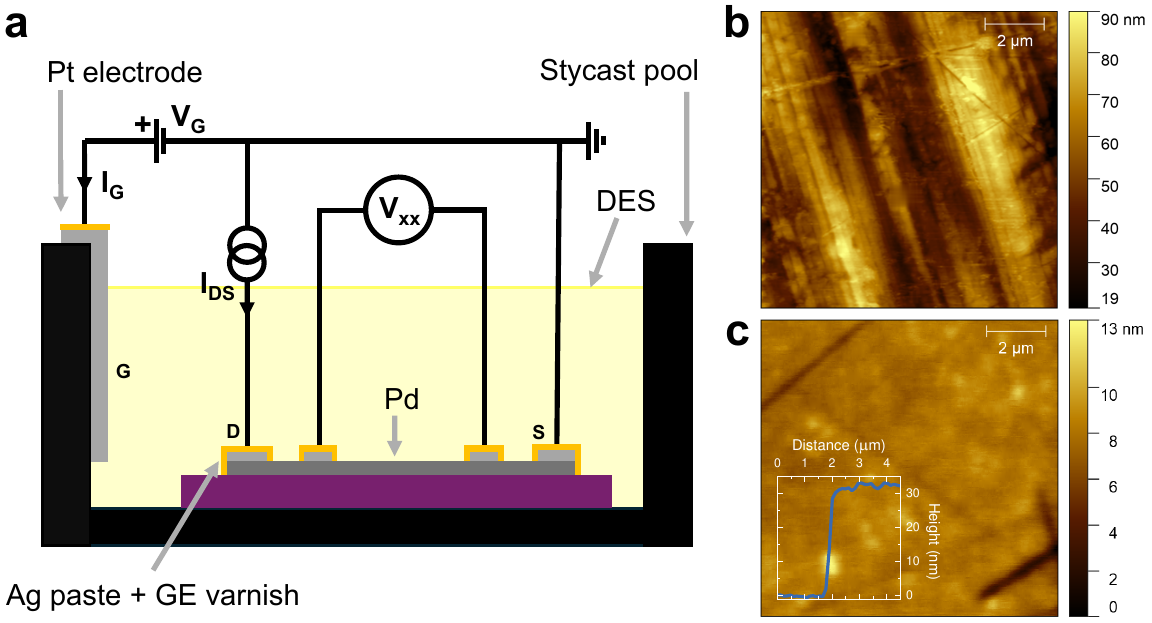}
  \caption{
  \textbf{Ionic gating-induced protonation: setup and materials.
  (a)} Schematic of the experimental setup designed to attain IGP within Pd samples while simultaneously monitoring their electrical resistivity in situ during the process.
  Source (S), drain (D), and Pt gate (G) electrodes are explicitly highlighted.
  Gate voltage $V\ped{G}$, gate current $I\ped{G}$, source-drain current $I\ped{DS}$, and four-wire voltage $V\ped{xx}$ are also indicated.
  \textbf{(b)} Representative AFM topography map of a Pd foil sample.
  \textbf{(c)} Representative AFM topography map of a Pd thin film sample and (inset) height profile across the film edge.
  }
  \label{fig:1}
\end{figure*}

Pd samples were electrically contacted by drop-casting silver paste (RS Components) onto thin platinum (Pt) wires.
The contacts were subsequently covered by a protective and electrically-insulating varnish (GE varnish).
The samples were were placed at the bottom of a custom-built Stycast pool (typical size $10\times5\times5$\,mm\apex{3}) equipped with a Pt counter electrode, which was then filled with the choline chloride-glycerol DES (as sketched in Fig.\,\ref{fig:1}a).
The gate voltage $V\ped{G}$ was applied to the Pt electrode under ambient conditions using the first channel of an Agilent B2912 source-measure unit (SMU) that also measured the gate current $I\ped{G}$.
The four-probe resistance was measured \textit{in situ} in the collinear configuration by sourcing a constant current $I\ped{ds}$ between drain (D) and source (S) contacts using the second channel of the same SMU (typical values $\approx10\,\upmu$A and $\approx1\,$mA for films and foils respectively), and measuring the longitudinal voltage drop $V\ped{xx}$ between inner voltage contacts with an Agilent 34420 nanovoltmeter.
The resistivity $\rho$ was calculated as $\rho = V\ped{xx}I\ped{ds}^{-1}\,twl^{-1}$, where $t$ and $w$ are the sample thickness and width, respectively, and $l$ is the distance between inner voltage contacts.
Common-mode offsets were eliminated by means of the current reversal method.
The entire setup was mounted on the cryogenic sample holders to allow for measurements of the resistance as a function of temperature immediately after the IGP process. 
Only for \textit{in situ} measurements in the \apex{3}He insert, due to the incompatibility of the Stycast pool with the insert dimensions, a standard sample holder was employed.
In this case, 
a planar configuration was used, with the Pt counterelectrode placed on the side of the Pd sample.
They were both covered by the same drop of DES, and a $15-\upmu$m polyimide foil was placed on top, to improve thermomechanical stability.
\subsection{Temperature-dependent electric transport measurements}\label{methods:rho-T}
The resistivity of the Pd samples was measured by directly cooling the entire IGP setup. 
Measurements were carried out as a function of temperature either in a Cryomech PT-403 pulse-tube cryocooler (base temperature $\approx2.8$\,K) or in the variable-temperature insert of an Oxford \apex{4}He cryostat (base temperature $\approx1.6$\,K). Ultralow temperature measurements were performed in a Cryogenic Ltd \apex{3}He insert (base temperature $\approx0.28$\,K) compatible with the Oxford cryostat.
Magnetic fields were applied perpendicular to the substrate plane using the 9\,T superconducting magnet of the Oxford cryostat.
\section{Results}

\subsection{Materials}
The measurements performed in this study were carried out on Pd samples in the form of bulk foils or thin films. 
The thickness of the commercial bulk foils was determined to be $25\pm1\,\upmu$m using a micrometer and their surface morphology was assessed via atomic force microscopy (AFM, see Section\,\ref{methods:AFM}).
A representative AFM topographic map of the foils is presented in Fig.\,\ref{fig:1}b.
The surface morphology is clearly dominated by the oriented scratch pattern that extends throughout the entire surface and that is likely due to the polishing process. 
The RMS and mean roughness were estimated to be approximately 14\,nm and 11\,nm, respectively.\\
\indent Thin Pd films were obtained by standard microfabrication techniques (see Section\,\ref{methods:samples}), and a representative AFM topographic map is presented in Fig.\,\ref{fig:1}c.
Again, no obvious granular structure could be detected and the films exhibited a significantly smoother surface compared to the foils, with RMS and mean roughness values of approximately 0.8\,nm and 0.5\,nm, respectively, attesting to the high quality of the deposited films.
The film thickness was determined to be $34\pm7$ nm by averaging over multiple step edges, one of which is displayed in the inset of Fig.\,\ref{fig:1}c.\\
\indent A choline chloride-glycerol DES was selected as the gate electrolyte owing to it readily providing dissociated H\apex{+} ions when cathodic voltages are applied to the working electrode\,\cite{Lee2016GC, Lin2017IJHE, Wang2023choline, Li_2016}, and synthesized as detailed in Section\,\ref{methods:DES}\,\cite{Mero2024GC}. For conciseness, this specific solvent will be referred to as DES throughout the manuscript.
\subsection{IGP process}
\begin{figure*}[t]
\centering\includegraphics[width=\textwidth, keepaspectratio]{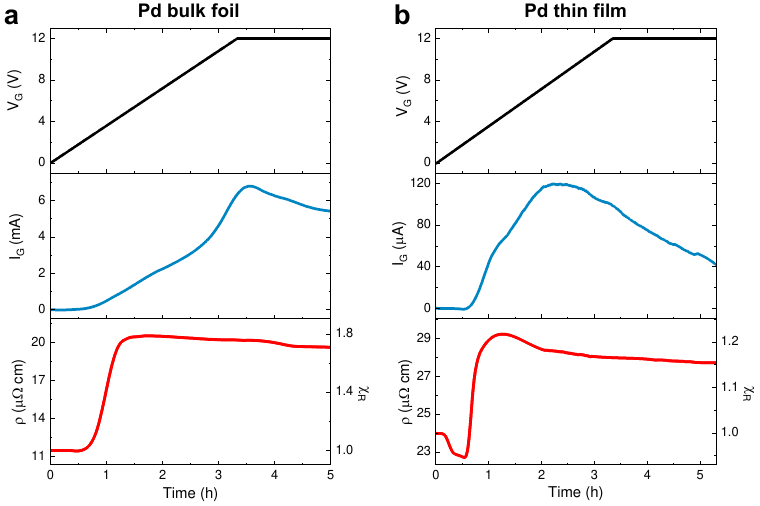}
  \caption{
  \textbf{Examples of the dynamics observed during ionic gating-induced protonation (IGP) of palladium (Pd) samples at room temperature.}
  \textbf{(a)} IGP of a bulk foil.
  \textbf{(b)} IGP of a thin film.
  Top panels display the applied gate voltage $V\ped{G}$, 
  middle panels display the measured gate current $I\ped{G}$, and 
  bottom panels display the electrical resistivity $\rho$ measured \textit{in situ} in absolute units (left scale) or normalized to its pristine value $\chi_\mathrm{R}=\rho(x)/\rho(0)$ (right scale).}
  \label{fig:2}
\end{figure*}
IGP was performed on both Pd foils and Pd films under ambient conditions implementing the experimental setup depicted in Fig.\,\ref{fig:1}a and further described in Section\,\ref{methods:IGP}.
The apparatus was integrated into a cryostat to enable rapid cooling upon completion of the gating process so as to minimize undesired H desorption from the Pd lattice. 
This configuration allowed for the application of a gate voltage $V\ped{G}$ across the electrolyte, between the Pt gate electrode and the sample itself.
Notably, voltages exceeding the established cathodic stability limit of the DES were applied to trigger the mechanisms responsible for the release of free H\apex{+} ions, which are prerequisites for H injection into Pd\,\cite{Wang2023choline, Li_2016, sinclair_2021}. A typical IGP run was carried out by ramping $V\ped{G}$ from 0\,V up to a specific final value -- exceeding the DES cathodic stability limit -- with a rate of 1\,mV/s.
This slow sweep rate ensured sufficient time for the ionic dynamics in the DES to adjust to the changing voltage.
The final value of $V\ped{G}$ was then held constant for a suitably long time as discussed below.\\
\indent To verify successful protonation, the setup allowed for simultaneous measurement of the gate current $I\ped{G}$ flowing between the gate (G) and the source (S) contact, and the in-situ resistivity $\rho$ of the sample undergoing treatment.
Monitoring $I\ped{G}$ provides insights into the response of the DES to voltage variations and of its dependence on the gating time.
However, it is crucial to note that not all generated H\apex{+} ions detected via $I\ped{G}$ necessarily contribute to protonation, since part of them can be captured in other electrochemical processes including hydrogen evolution reaction\,\cite{PdHERreview}.
Therefore, monitoring $\rho$ is essential to confirm H\apex{+} injection and its impact on modifying the electronic properties of Pd.
Representative examples of the observed in-situ IGP dynamics are presented in Fig.\,\ref{fig:2} where the time evolution of the three key quantities $V\ped{G}$, $I\ped{G}$, and $\rho$ is displayed in the case of gated bulk foils (panel a) and thin films (panel b).
The dimensionless quantity $\chi_\mathrm{R} = \rho(x)/\rho(0)$ represents the ratio between the resistivity of the sample being protonated (at a certain H/Pd ratio $x$) and its resistivity in the pristine state.\\
\indent In Fig.\,\ref{fig:2}, all $I\ped{G}$ profiles exhibit very low values (few nA) for applied $V\ped{G}\lesssim1.9$\,V.
This value is somewhat larger than the cathodic stability limit established for the DES under investigation with glassy carbon working electrodes (1.38\,V)\,\cite{Wang2023choline, Li_2016, sinclair_2021}. 
However, the specific oxidation potential of glycerol is strongly related to electrode material and can in principle acquire values even below 1\,V vs reference electrode. For example, Lee et al.\,\cite{Lee2016GC} reported the oxidation of glycerol at 
0.80\,V (vs SHE, standard hydrogen electrode) over a PtSb/C catalyst; whereas Lin et al.\,\cite{Lin2017IJHE} performed the oxidation of glycerol with NiNPs/ITO electrode in the 0.2\,M NaOH operating at a potential of 0.70\,V vs Ag/AgCl.
In our setup, the electrochemical processes responsible for H\apex{+} release remain inactive within the range $V\ped{G}\lesssim1.9$\,V, resulting in a negligible $I\ped{G}$.
However, reaching and exceeding $V\ped{G}\approx1.9\,$V triggers a significant rise in $I\ped{G}$, signaling the onset of the electrochemical reactions responsible for the release of H\apex{+} ions.
Further increasing $V\ped{G}$ beyond these values initially increases $I\ped{G}$ even further, as the H\apex{+} ion release is strengthened, but eventually leads to $I\ped{G}$ reaching a maximum and then decreasing. 
Part of this decrease can be attributed to the pseudocapacitive dynamics of the polarized DES\,\cite{ScholtzBook} since its onset is roughly correlated with the time scale at which the $V\ped{G}$ ramp reaches its final value;
however, at least part of the decrease is likely associated with the progressive degradation of the electrolyte resulting from the application of voltages exceeding the DES stability window, since such conditions would eventually hamper the capability of the electrolyte to sustain a stable production of free H\apex{+} ions.
Overall, $V\ped{G}=12$\,V was selected as the maximum applicable value in the gate ramps since rapid device degradation and failure systematically occurred upon the application of even larger voltages.\\
\indent A more quantitative assessment of the impact of the application of $V\ped{G}$ to the Pd samples can be obtained by monitoring $\rho$ or better the ratio $\chi_\mathrm{R}$.
Previous studies have demonstrated that $\chi_\mathrm{R}$ for PdH$_x$ increases monotonically with H content, reaching a peak at $x\approx0.7$, and then decreases monotonically, eventually returning to unity at $x=1$\,\cite{Zhang2002JEC, Flangan_Lewis_1961, Smith_Otterson_1970, Barton_Lewis_1962}.
Notably, in our case, both Pd foils and Pd films exhibit analogous $\chi_\mathrm{R}$ profiles over time, suggesting a direct correlation between gating time and H uptake.
As illustrated in Fig.\,\ref{fig:2}, $\chi_\mathrm{R}$ typically increases to a peak and then gradually decreases towards a constant value.
This stabilization can be attributed to a dynamic equilibrium between H intrinsically desorbed by Pd and H injected from the electrolyte.\\
%
\indent Although gated Pd films and foils exhibit similar overall trends in their time-dependent $\chi_\mathrm{R}$ curves, the reduced thickness of the thin films introduces distinct features compared to bulk foils.
First, the overall variation in $\chi_\mathrm{R}$ is significantly lower for Pd films ($\chi_\mathrm{R}$ up to $\approx1.4$) as compared to foils ($\chi_\mathrm{R}$ up to $\approx1.8$), an observation that aligns with previous studies on Pd thin films\,\cite{Wagner_2010, Lee_Glosser_1986} and that was attributed mainly to a stress-dependent reduction of H solubility at a given H pressure\,\cite{Wagner_2010}.
Second, a localized decrease in $\rho$ (and consequently in $\chi_\mathrm{R}$) is observed at low voltages ($V\ped{G}\lesssim1.9$\,V) in the thin films (Fig.\,\ref{fig:2}b), at variance with the almost constant $\rho$ exhibited by bulk foils (Fig.\,\ref{fig:2}a) in the same voltage range.
Since in this range the electrochemical processes for H production are inactive, this distinct feature observed only in the thin films can be attributed to an electrostatic gating effect.
In this low-voltage regime, the primary effect of the applied $V\ped{G}$ is to accumulate charges at the Pd-electrolyte interface, resulting in an effective increase in carrier concentration and, thus, a reduction in resistivity.
This behavior agrees with well-established observations in ion-gated metallic films\,\cite{Daghero_2012, Tortello_2013, Li_2016_control, Piatti2019PRM, Xi2016PRL, Piatti2021NE, Dushenko2018NatCommun, Piatti2022PRApp}.
It is noteworthy that this effect occurs in both Pd foils and films but leads to a non-negligible change in $\rho$ only in the latter due to their much smaller thickness (34\,nm vs 25\,$\upmu$m) enabling the detection of the surface-bound electrostatically-modulated layer\,\cite{Daghero_2012, Piatti2021NE, Piatti2022PRApp, Piatti2017PRB, Piatti2019EPJ, Piatti2020ApSuSc}.

\subsection{$\rho-T$ measurements}
\begin{figure*}[t]
    \centering\includegraphics[width=\textwidth, keepaspectratio]{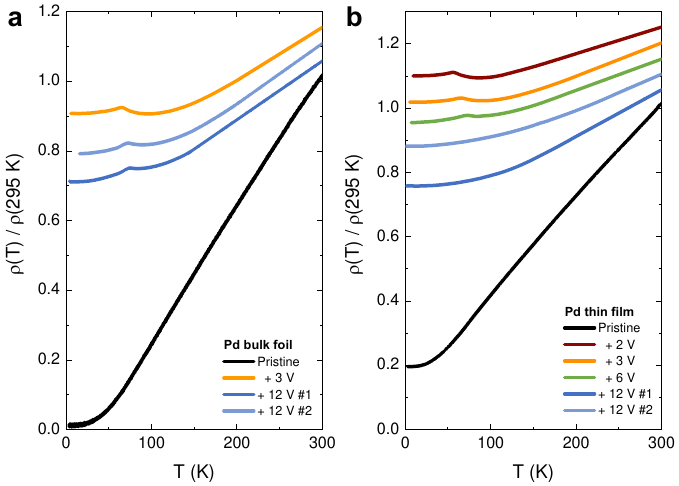}
  \caption{
  \textbf{Temperature dependence of the resistivity of PdH$_x$ samples before ($x=0$) and after ($x>0$) ionic-gating induced protonation.}
  \textbf{(a)} $\rho-T$ of bulk foils.
  \textbf{(b)} $\rho-T$ of thin films.
  All curves are normalized to their value at $T=295$\,K and labelled according to the final value of the applied gate voltage.
  Normalized curves at $V\ped{G}>0$ are vertically shifted by 0.05 for clarity.
  }
  \label{fig:3}
\end{figure*}
To further verify that the observed $\chi_\mathrm{R}$ changes for $V\ped{G}\gtrsim1.9$\,V are due to successful protonation, we measured the dependence of $\rho$ on the temperature $T$ following each IGP process, in both bulk foils (Fig.\,\ref{fig:3}a) and thin films (Fig.\,\ref{fig:3}b) as detailed in Section\,\ref{methods:rho-T}.\\
\indent In the case of Pd bulk foils, the pristine $\rho(0)$ was found to smoothly decrease as $T$ was reduced and then saturate for $T\lesssim5$\,K (solid black line in Fig.\,\ref{fig:3}a), as expected for a typical metallic material\,\cite{Matula_1979}.
The value of $\rho(0)$ measured at 295\,K was $11.5\,\upmu\Omega$\,cm (as reported also in Fig.\,\ref{fig:2}a) and is in agreement with reported values for bulk Pd\,\cite{Baba_1990, Matula_1979}.
In contrast, the $T$ dependences of $\rho$ in the gated foils (solid colored lines in Fig.\,\ref{fig:3}a) exhibit significant deviations from the pristine state, indicating successful H incorporation into the Pd lattice.
Notably, a peak emerges in the $\rho-T$ curves around 50\,K, corresponding to the well-documented anomaly observed in various physical properties of PdH$_x$\,\cite{Skoskiewicz_1968, Zepeda1971}, including specific heat\,\cite{Mitacek_1963, Nace1957}, internal friction\,\cite{Jacobs_1976}, and thermal relaxation\,\cite{Jacobs_1977}.
This feature was initially attributed to an order-disorder phase transition\,\cite{Manchester_1994, Blaschko_1984_1}.
However, current understanding attributes this anomaly to a kinetic effect arising from changes in the short-range order of H\,\cite{SETAYANDEH_2020, Akiba_2015}.
Here, the temperature corresponding to this anomaly ($T\ped{a}$) is defined as the local maximum of the $\rho-T$ curve in the range $0-100$\,K, as further discussed in Section\,\ref{sec:discussion_T_anomaly}.\\
\indent The detection of this anomaly is crucial because the value of $T\ped{a}$ is known to correlate with the H concentration in PdH$_x$. Namely, for $x >0.6$, $T\ped{a}$ is expected to increase as a function of $x$, while the peak height decreases, and eventually vanishes at higher concentrations\,\cite{Ho_1968, HARPER_1974}.
Such an increase in $T\ped{a}$ can indeed be observed in the $\rho-T$ curves of our gated foils when $V\ped{G}$ is increased from 3 to 12\,V (solid orange line and solid blue lines in Fig.\,\ref{fig:3}a respectively), and therefore confirms that the decrease in $\chi_\mathrm{R}$ shown in Fig.\,\ref{fig:2}a for $V\ped{G}\gtrsim 5$\,V is primarily due to increasing H uptake rather than desorption.
However, the persistence of the anomaly in the $\rho-T$ curves measured even in the foils gated at $V\ped{G}=12$\,V suggests that, in this case, the H concentration does not reach the threshold required for a detectable superconducting transition, since the resistivity anomaly is not observed in PdH$_x$ samples with optimal critical temperatures\,\cite{Standley1979SSC, Schirber_1974}.
Consistently, we observed no trace of a superconducting transition in Pd foils gated by applying a final $V\ped{G}$ up to the maximum value of 12\,V.\\ 
\indent In the case of the Pd thin films, $\rho(0)$ also exhibited a typical metallic behavior (solid black line in Fig.\ref{fig:3}b) and its value measured at 295\,K ($24\,\upmu\Omega$\,cm, Fig.\,\ref{fig:2}b) was consistent with previous results for 35\,nm-thick Pd samples\,\cite{De_1988, Satrapinski_2011}.
Similar to foils, the application of $V\ped{G}\gtrsim1.9\,$V induced a significant deviation of the $\rho-T$ curves from the pristine behavior, indicating successful H incorporation into the Pd lattice.
The anomaly is evident for final values of $V\ped{G}$ between 2\,V and 6\,V (solid brown, orange and green curves in Fig.\,\ref{fig:3}b), with a progressive shift towards higher temperatures indicating an increase in injected H.
Also in this case, this observation further supports our interpretation that the decrease in $\chi_\mathrm{R}$ shown in Fig.\,\ref{fig:2}b is due to H uptake rather than desorption.\\
\begin{figure}[t]
    \centering\includegraphics[width=\columnwidth, keepaspectratio]{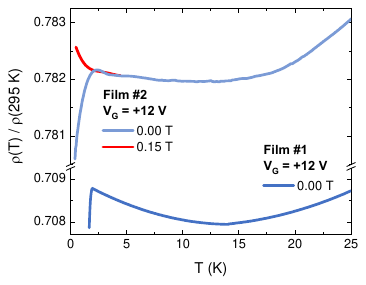}
  \caption{
  \textbf{Partial superconducting transitions in gated Pd thin films.}
  Temperature dependence of the resistivity (normalized to its value at 295\,K) attained by applying a final $V\ped{G}=12$\,V for $\approx2$ hours to two Pd thin films.
  Film \#1 (dark blue line) was measured in the $^4$He cryostat down to 1.67\,K in a zero applied magnetic field.
  Film \#2 (light blue and red lines) was measured in the $^3$He cryostat down to 0.35\,K in an applied magnetic field of 0\,T and 0.15\,T respectively.
  The curves at 0\,T are an expanded view of the same data shown in Fig.\,\ref{fig:3}b.
  }
  \label{fig:4}
\end{figure}
\indent In contrast with bulk foils, however, when a $V\ped{G}= 12$\,V is applied to the thin films the anomaly peak disappears and the $\rho-T$ curves (solid blue lines in Fig.\,\ref{fig:3}b) become again smooth and featureless, albeit with a much larger residual resistivity with respect to the pristine $\rho(0)$.
This suggests that, unlike in the bulk foils, in the thin films the H concentration can be driven beyond the limit where the anomaly is expected to vanish ($x \sim 0.85$)\,\cite{Manchester_1994}; the reason for this discrepancy is unclear, but we speculate that it might be caused by the larger surface-to-volume ratio of the films with respect to the foils, which promotes an increase in the maximum H uptake under comparable gating conditions.
Thanks to this, a finite superconducting transition temperature should become observable, and indeed the $\rho-T$ curves measured at $V\ped{G}=12$\,V in thin films do exhibit very small drops for $T\lesssim2$\,K, which can be interpreted as a signature of superconducting transitions\,\cite{Skoskiewicz1972, Skośkiewicz_1973, SCHIRBER_1973, Stritzker1972, HARPER_1974, OCHMANN_1983, Krahn_1978}.
An illustrative example of this feature is given in Fig.\,\ref{fig:4}, which provides an expanded view of the curves displayed in Fig.\,\ref{fig:3}b for the two films gated at $V\ped{G}=+12$\,V -- the first measured in a $^4$He cryostat down to 1.67\,K (solid dark blue line) and the second in a $^3$He cryostat down to 0.35\,K (solid light blue line); see Section\,\ref{methods:rho-T} for details.
The superconducting nature of these resistivity drops is corroborated by the fact that the drop completely disappears upon applying a magnetic field of 150\,mT (solid red line in Fig\,\ref{fig:4}).
This field value was deliberately selected to exceed the zero-temperature upper critical magnetic field of PdH$_1$, $H\ped{c2}(0)\approx110$\,mT \cite{Skośkiewicz_1973, Horobiowski_1977, Sansores_1981}.
However, to definitively determine why only a partial transition is occurring down to 0.35\,K, a quantitative analysis of the H concentration is necessary.
This will clarify whether the absence of a full transition is due to an insufficient overall concentration or to an inhomogeneous H distribution within the film. 

\section{Discussion}

\begin{figure*}[t]
  \centering\includegraphics[width=\textwidth, keepaspectratio]{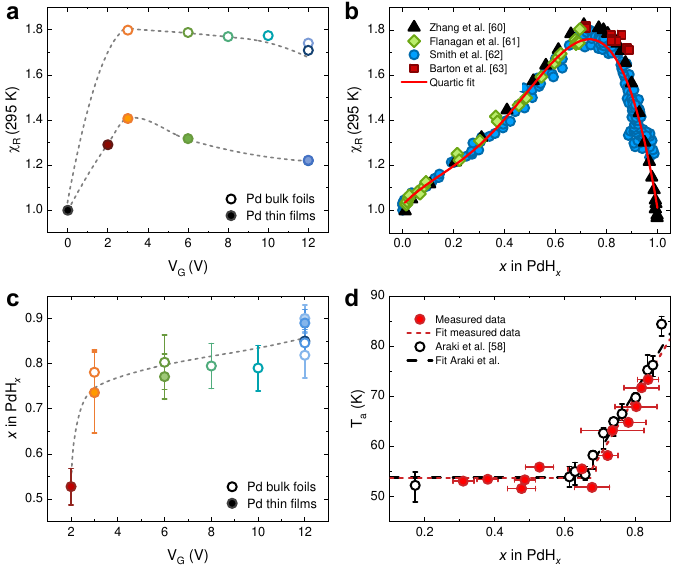}
  \caption{
  \textbf{Determination of the hydrogen concentration in gated Pd samples.}
  \textbf{(a)} Resistivity ratio $\chi_\mathrm{R}$ as a function of the final applied gate voltage $V\ped{G}$ for both bulk foils (hollow circles) and thin films (filled circles) after each IGP process. The adopted color code matches that used in Fig.\,\ref{fig:3}. Grey dashed lines are guides to the eye.
  \textbf{(b)} Empirical correlation between $\chi_\mathrm{R}$ and the hydrogen (H) content $x$ based on key studies from the literature. Symbols are the data adapted from Refs.\,\cite{Zhang2002JEC, Flangan_Lewis_1961, Smith_Otterson_1970, Barton_Lewis_1962}. The solid red curve is the fit to Eq.\,\eqref{model_Zhang2002}. 
  \textbf{(c)} $V\ped{G}$ dependence of $x$ for gated bulk foils (hollow circles) and thin films (filled circles) indirectly derived from the $\chi_\mathrm{R}$ values shown in panel (a). Grey dashed line is a guide to the eye.
  \textbf{(d)} Characteristic temperature, $T\ped{a}$, of the anomaly observed in the $\rho-T$ curves of PdH$_x$ as a function of $x$. Filled red circles are the data obtained in the present study, hollow black circles are the data adapted from Ref.\,\cite{Araki_2004}. Dashed lines are linear fits to the two data sets as discussed in the main text.
  }
  \label{fig:5}
\end{figure*}

\subsection{Determination of the H concentration}
A quantitative evaluation of the H concentration for each IGP process is crucial to elucidate the origin of the observed partial superconducting transition and to assess the performance of the electrolyte in facilitating H loading in Pd.
Direct measurements of the H concentration, such as $^1$H nuclear magnetic resonance\,\cite{Cui_2019, Piatti2023CommunPhys, Prando2023PRM}, require dedicated experimental setups and the use of bulk samples, making them ill-suited for \textit{in situ} measurements on thin films.
As a consequence, we employ an indirect estimation method where the H concentration is inferred from the resistivity ratio $\chi_\mathrm{R}$, which is known to correlate with $x$\,\cite{Zhang2002JEC, Flangan_Lewis_1961, Smith_Otterson_1970, Barton_Lewis_1962}.
Fig.\,\ref{fig:5}a presents the final measured $\chi_\mathrm{R}$ as a function of the final applied $V\ped{G}$ for IGP on both Pd bulk foils (hollow circles) and Pd thin films (filled circles).
This behavior reflects the time evolution of $\chi_\mathrm{R}$ displayed in Fig.\,\ref{fig:2} and is similarly attributed to a rising H concentration with increasing $V\ped{G}$.
To validate this attribution, we exploit previous studies that have quantified the correlation between $\chi_\mathrm{R} $ and $x$. This enables us to associate an $x$ content to each value of $\chi_\mathrm{R}$.
Fig.\,\ref{fig:5}b summarizes the $\chi_\mathrm{R}-x$ data obtained from a selection of key studies (symbols) that demonstrate this correlation for bulk Pd\,\cite{Zhang2002JEC, Flangan_Lewis_1961, Smith_Otterson_1970, Barton_Lewis_1962}.
It is important to note that the data from Ref.\,\cite{Smith_Otterson_1970} were measured at 273\,K, while all other datasets were acquired at 295\,K.
To ensure consistency across all studies, a temperature correction procedure, detailed within Appendix\,\ref{appendix:flanagan_correction}, is applied to this specific dataset.
Following a similar approach to Zhang et al.\,\cite{Zhang2002JEC}, a fourth-order polynomial fit is performed on the data presented in Fig.\,\ref{fig:5}b.
This fitting procedure yields the following empirical relationship between $\chi_\mathrm{R}$ and $x$: 
\begin{equation}
\chi_\mathrm{R}= a + bx+cx^2+dx^3+ex^4 \label{model_Zhang2002}
\end{equation}
where $a=(1.023 \pm 0.015)$, $b=(1.2 \pm 0.3)$, $c=(-3.3 \pm 1.1)$, $d=(9.9 \pm 1.7)$ and $e=(-7.7 \pm 0.8)$. This equation is shown as the solid red curve in Fig.\,\ref{fig:5}b.
This established relationship allows us to solve for the unknown H concentration $x$ associated with the measured $\chi_\mathrm{R}$ values for bulk Pd.\\
\indent However, before directly solving this equation, an adjustment is necessary to account for the thickness-induced deviation of the $\chi_\mathrm{R}-x$ relationship from the bulk behavior exhibited by our thin-film data (hollow circles in Fig.\,\ref{fig:5}a)\,\cite{Wagner_2010, Lee_Glosser_1986}.
Direct calculation of $\chi_\mathrm{R}$ without thickness adjustment would lead to erroneous estimations of H concentration.
To address this issue, we follow Refs.\,\cite{Wagner_2010, Lee_Glosser_1986} that showed that the reduced thickness of Pd thin films primarily introduces a uniform suppression in the $\chi_\mathrm{R}-x$ curve compared to Pd bulk samples, without altering its overall shape.
To account for this effect in the determination of $x$ in our gated thin films, we employ a linear transformation to map the measured film $\chi_{\mathrm{R},\,\text{film}}$ onto the desired $\chi_{\mathrm{R},\,\text{bulk}}$.
This transformation can be expressed as: 
\begin{equation}
\chi_{\mathrm{R},\,\text{bulk}}(x)= m \cdot \chi_{\mathrm{R},\,\text{film}}(x)+ q
\label{eq:chi_film_rescaling}
\end{equation}
where the parameters $m$ and $q$ are determined to preserve the intrinsic shape of the curve.\\
A first condition on $m$ and $q$ is given by the definition of $\chi_\mathrm{R}$, which at $x=0$ gives $\chi_{\mathrm{R},\,\text{bulk}}(0) = \chi_{\mathrm{R},\,\text{film}}(0)=1$ and thus:
\begin{equation}
    m + q = 1
\label{eq:chi_film_zero_rescaling}
\end{equation}
A second condition on $m$ and $q$ is obtained by ensuring that the peak position remains unaffected by the transformation, which holds if:
\begin{equation}
\chi_{\mathrm{R},\text{\,bulk,\,max}} = m \cdot \chi_{\mathrm{R},\text{\,film,\,max}} + q
\label{eq:chi_film_max_rescaling}
\end{equation}
where $\chi_{\mathrm{R},\text{\,bulk,\,max}}$ is the maximum value obtained from Eq.\,\eqref{model_Zhang2002} and $\chi_{\mathrm{R},\text{\,film,\,max}}$ is the maximum $\chi_\mathrm{R}$ value derived from each specific IGP process (see Fig.\,\ref{fig:2}b).\\
Combining Eq.\,\eqref{eq:chi_film_zero_rescaling} and Eq.\,\eqref{eq:chi_film_max_rescaling} finally yields:
\begin{equation}
m = \frac{1 - \chi_{\mathrm{R},\,\text{bulk,\,max}}}{1 - \chi_{\mathrm{R},\text{\,film,\,max}}} 
\end{equation}
which in turn enables the subsequent application of Eq.\,\eqref{model_Zhang2002}.\\
\indent It is necessary to point out that the polynomial Eq.\,\eqref{model_Zhang2002} yields two distinct real solutions for $x$ at a given $\chi_\mathrm{R}$ value, with the exception of the maximum point of the curve.
To discriminate between these two solutions and ensure a unique mapping, we consider the time evolution of $\rho$ during each IGP process, as exemplified in Fig.\,\ref{fig:2}: Specifically, we determine whether each given $\chi_\mathrm{R}$ value was attained on the ascending or descending branch of each $\chi_\mathrm{R}$ vs time curve.
%
%
Knowing the position relative to the maximum point allows us to discard one of the two solutions derived from the equation, thereby achieving a one-to-one correspondence between $\chi_\mathrm{R}$ and $x$. 
The uncertainty on this procedure was determined from the inherent variability observed in the literature data used to establish the equation linking $\chi_\mathrm{R}$ with $x$ (as displayed in Fig.\,\ref{fig:5}b).
Therefore, for each value of $\chi_\mathrm{R}$, we calculated the semi-difference of the $x$ spread of the data points presented in Fig.\,\ref{fig:5}b.
This semi-difference was subsequently assigned as the uncertainty of the $x$ value obtained by solving Eq.\,\eqref{model_Zhang2002} for the given value of $\chi_\mathrm{R}$.
The results obtained using this indirect approach for the determination of the H concentration are presented in Fig.\,\ref{fig:5}c, plotted as a function of the final value of $V\ped{G}$ reached during each IGP process.

\subsection{Validation of the indirect H estimation}\label{sec:discussion_T_anomaly}
The indirect method employed for the estimation of the H concentration using the $\chi_\mathrm{R} - x$ correlation is susceptible to systematic errors. To ensure the validity of these results, an independent validation is crucial.
To this end, we exploit the anomaly observed in the $\rho-T$ curves of PdH$_x$, since the temperature $T\ped{a}$ at which it appears monotonically increases with increasing H concentration for $x >0.6$\,\cite{Ho_1968, HARPER_1974, Araki_2004, Manchester_1994}.
For each $\rho-T$ curve measured in our gated Pd samples, we independently determine the corresponding value of $T\ped{a}$ as the position of the local maximum and the value of $x$ using the previously-described method based on $\chi_\mathrm{R}$.
Specifically, $T\ped{a}$ was determined by a parabolic fitting procedure:
For samples with H concentrations $0.5\leq x \leq0.8$, each $\rho-T$ curve exhibited a clear local maximum within the $40-100$\,K temperature range so that a parabolic function of the form $\rho(T)=a+b \cdot (T - T\ped{max})^2$ could be directly fitted to the data surrounding the local maximum.
For samples with H concentrations $x < 0.5$ or $x > 0.8$, a discernible anomaly persisted within the $\rho-T$ curves but deviated from a simple local maximum, manifesting instead as a pronounced shoulder superimposed to the overall metallic trend.
In these curves, a background subtraction was employed to effectively remove the underlying metallic contribution.
This process transformed the shoulder into a more distinct peak, which was subsequently analyzed using the previously described parabolic fitting procedure.
In all cases, the parameter $T\ped{max}$, along with its associated uncertainty, was then taken as the estimated value of $T\ped{a}$.\\
\indent Fig.\,\ref{fig:5}d presents a comparison between our resulting $T\ped{a}-x$ data points (filled red circles) and the data extracted from a representative study by Araki et al.\,\cite{Araki_2004} (hollow black circles). 
An excellent agreement is observed between our findings and those of Ref.\,\cite{Araki_2004} where $T\ped{a}$ was determined to remain constant upon increasing $x$ up to the concentration marking the phase boundary between so-called $\alpha+\beta$ and pure $\beta$ phases of PdH$_x$, followed by a linear increase in the pure $\beta$ phase\,\cite{Manchester_1994}.

Fitting the data of Ref.\,\cite{Araki_2004} to this model (dashed black lines in Fig.\,\ref{fig:5}d) yields a constant temperature of $53.8\pm1.2$\,K up to $x = 0.65$ and $T\ped{a} = -22 \pm 6\,\mathrm{K} + x(116 \pm 7\,\mathrm{K})$ for $x > 0.66$.
Assuming a similar behavior for our system and repeating the fits (dashed red lines in Fig.\,\ref{fig:5}d), we observe a constant temperature of $53.8 \pm 1.6$\,K up to $x = 0.69$, while the relationship 
$T\ped{a} = -23 \pm 10\,\mathrm{K} + x(116 \pm 14\,\mathrm{K})$ well approximates our data for $x > 0.69$.
This finding is consistent with Ref.\,\cite{Araki_2004} within the reported uncertainties, confirming the reliability of the indirect method employed for the determination of the H concentration in our gated Pd samples.

\subsection{Evaluation of the DES performance} 
The robustness of the methods we employed for the determination of the H concentration in our gated Pd samples allows for a reliable interpretation of the trend observed in Fig.\,\ref{fig:5}c.
A clear positive correlation is evident between the concentration of injected H and the gating voltage exceeding the cathodic limit of the DES. Notably, the H concentration exhibits a sharp rise at low $V\ped{G}\sim 2-3$\,V, followed by a continuous but slower increase at higher $V\ped{G} > 3$\,V.
This efficient H loading without apparent damage to the Pd sample is observed up to a maximum $V\ped{G}=12$\,V, above which rapid device degradation sets in.
Upon the application of this maximum $V\ped{G}$ for approximately 2 hours, a maximum H concentration $x = 0.85 \pm 0.03$ was obtained in the Pd bulk foils.
Downscaling the volume of the Pd sample allowed the further enhancement of the H concentration up to $x = 0.89 \pm 0.03$ in the Pd thin films.
%
%
\begin{table*}
    \centering
    \begin{tabular}{cccccc}
    \toprule
        Compound & Electrolyte & Stoichiometry & Cell volume & H concentration & Ref. \\
         &  &  & {\r{A}}$^3$ & H cm\apex{-3} &  \\
    \midrule
       Pd  & Choline chloride:glycerol 1:3 & PdH\ped{0.89} & 16.85\,\cite{Schirber_1975} & $5.3\times 10^{22}$ & This work \\
       Pd  & 1 N HCl solution & PdH\ped{0.70} & 16.46\,\cite{Schirber_1975} & $4.2\times 10^{22}$ & \cite{Flangan_Lewis_1961} \\
       Pd  & 1 N HCl solution & PdH\ped{0.89} & 16.85\,\cite{Schirber_1975} & $5.3\times 10^{22}$ & \cite{Barton_Lewis_1962} \\
       Pd  & 
       5.25 mol/l H$_2$SO$_4$ solution & PdH$_{0.88}$ & 16.89\,\cite{Schirber_1975}  & $5.2\times 10^{22}$ & \cite{Araki_2004} \\
       Pd  & 0.1 N H$_2$SO$_4$ solution & PdH$_{0.99}$ & 16.98\,\cite{Schirber_1975} & $5.8\times 10^{22}$ & \cite{Smith_Otterson_1970} \\
       Co & GdO$_x$ solid-state electrolyte & CoH$_{0.1}$ & 11.14\,\cite{Fernandez1987} & $0.9\times 10^{22}$ & \cite{Tan_2019} \\
       WO$_3$ & DEME-TFSI ionic liquid & H$_{0.35}$WO$_3$ & 110\,\cite{Wang_2017} & $0.3\times10^{22}$ & \cite{Wang_2017}  \\
       SrRuO$_{3}$ & Ionic liquid (not specified)  & H$_1$SrRuO$_3$  &  62.5\,\cite{Li_2020} & $1.6\times10^{22}$ & \cite{Li_2020} \\
       SrFeO$_{2.5}$ & HMIM-TFSI ionic liquid & H$_{1}$SrFeO$_{2.5}$ & 60.3\,\cite{Yan_2024} & $1.7\times10^{22}$ & \cite{Yan_2024} \\
       SrCoO$_{2.5}$  & EMIM-BF$_4$ ionic liquid & H$_{1.14}$SrCoO$_{2.5}$ & 61.53\,\cite{Lu2017} & $1.8\times10^{22}$ & \cite{Lu2017} \\
       SrCoO$_{2.5}$  & DEME-TFSI ionic liquid & H$_{2.2}$SrCoO$_{2.5}$ & 65.1\,\cite{Li_2019ADVSCI} & $3.1\times10^{22}$ & \cite{Li_2019ADVSCI}\\
       NiCo$_2$O$_4$  & DEME-TFSI ionic liquid & H$_{1.86}$NiCo$_2$O$_4$ & 71.39\,\cite{Wang_2019} & $2.6\times10^{22}$ & \cite{Wang_2019} \\
       TiSe\ped{2}  & EMIM-BF$_4$ ionic liquid & H$_2$TiSe$_2$ & 65.7\,\cite{Piatti2023CommunPhys} & $3.0\times10^{22}$ & \cite{Piatti2023CommunPhys} \\
    \bottomrule
    \end{tabular}
    \caption{{\color{blue}Summary of the maximum H concentrations attained in different compounds via electrochemical loading. Where not already explicitly stated in the corresponding references, the concentrations have been obtained by dividing the stoichiometries by the cell volumes. All cell volumes are reported per formula unit.}}
    \label{tab:H_concentrations}
\end{table*}
\add{We now compare the efficiency of our DES-based protonation method with a selection of results from the literature as summarized in Table\,\ref{tab:H_concentrations}, which reports the maximum H concentration attained in different materials via electrochemical loading.
Taking into account the loading-dependent unit cell volume of PdH$_x$\,\cite{Schirber_1974}, our method attains a maximum H concentration of $\approx 5.3\times 10^{22}$\,H\,cm$^{-3}$.
When compared to existing electrochemical synthesis routes for PdH$_x$, this value is comparable with the best results obtained exploiting HCl-based electrolytes ($\approx 5.3\times 10^{22}$\,H\,cm$^{-3}$)\,\cite{Flangan_Lewis_1961, Barton_Lewis_1962} and reaches more than $\approx 91\%$ of those exploiting H$_2$SO$_4$-based electrolytes ($\approx 5.8\times 10^{22}$\,H\,cm$^{-3}$)\,\cite{Araki_2004, Smith_Otterson_1970}.
When one considers H loading attained via gate-driven protonation with ionic liquids on other materials, our method demonstrates a $\approx 70\%$ improvement over the best values attained both on protonated oxides ($\approx 3.1\times 10^{22}$\,H\,cm$^{-3}$)\,\cite{Tan_2019, Wang_2017, Li_2020, Yan_2024, Lu2017, Li_2019ADVSCI, Wang_2019} and transition metal dichalcogenides ($\approx 3.0\times 10^{22}$\,H\,cm$^{-3}$)\,\cite{Piatti2023CommunPhys}.\\
}
\indent Based on previous studies, we expect PdH$_{0.89 \pm 0.03}$ to exhibit a superconducting transition with a mid-point critical temperature between 1.9\,K and 4.3\,K\,\cite{Standley1979SSC}.
However, this behavior is not observed in our experiment: As shown in Fig.\,\ref{fig:4}, although we detect the onset of a superconducting transition at 1.9\,K (defined as when the $\rho-T$ curve at zero field appreciably deviates from the one measured at $B=150$\,mT), a clear mid-point is absent even at temperatures as low as 0.35\,K.
We attribute this discrepancy to significant inhomogeneity in the H concentration within the Pd film, which likely broadens the transition.
\add{
In particular, secondary-ion mass spectroscopy measurements in protonated oxide films\,\cite{Wang_2017, Rafique2019Nano, Ri_2020, Li_2020, Shen_2021} showed that the gate-driven H concentration may be peaked close to the film surface and then decrease deeper within the bulk.
If such a distribution were to occur in our Pd films, the superconducting state might be attained only in a thin surface layer and might be non-percolating, thereby resulting in the observed incomplete resistive transition.
Developing a method to improve the homogeneity of the H concentration would therefore be highly beneficial towards the synthesis of high-quality samples of H-based superconductors and other functional materials.
In this context, we envisage that the primary parameter to optimize is the ionic diffusivity during the gate-driven protonation, within both the gate electrolyte and the protonated material. 
The former can be improved by decreasing the viscosity of the electrolyte and increasing its ionic conductivity\,\cite{Mero2023}; the latter chiefly by increasing the temperature at which the gate-driven protonation process is carried out\,\cite{Meng2022PRB, Wang_2019, Wu2023, Wu_2024ACS, Piatti2018ApSuSc_MoS2}.
However, care must be taken since other electrochemical processes -- including electrolyte dissociation and potential etching of the gated material\,\cite{Lu2017} -- would be promoted as well by the temperature increase, which might require the design and usage of DESs suitable to operate under such conditions.\\
\indent At this end, we can outline the following guidelines for the design of a DES with improved performances for the synthesis of H-rich materials via gate-driven protonation:}
{\color{blue}
\begin{enumerate}
    \item A high density of protons is available for gate-driven injection.
    \item A low viscosity to attain a large ionic conductivity and improve the homogeneity of the H transfer from the electrolyte to the material.
    \item A low threshold voltage for the release of free H, so that the gate voltage can be kept low to minimize damage to sensitive materials.
    \item A large electrochemical window against electrochemical processes other than the release of H, to minimize undesired electrochemical interactions with the material and enable the application of large voltages to stabilize phases with high H concentrations.
    \item A wide temperature range where the DES is liquid, both above and below room temperature, to allow for fine tuning of the protonation process by optimizing the processing temperature. 
\end{enumerate}}

\section{Conclusions}
\noindent In this work, we demonstrated the feasibility of employing a deep eutectic solvent choline chloride:glycerol 1:3 as a gate electrolyte for efficient H loading into Pd via ionic gating-induced protonation.
The effect of the applied gate voltage on H uptake was evaluated by monitoring changes in the electrical resistivity of the gated Pd samples.
A custom analysis method was employed to indirectly quantify the H concentration attained following each protonation process, evidencing a positive correlation between applied gate voltage and H uptake within the Pd lattice.
This trend was evident in both Pd foils and films. Notably, at the maximum gate voltage of 12\,V explored in this study, this straightforward technique resulted in H concentrations potentially sufficient for superconductivity to emerge.
Nevertheless, only partial superconducting transitions were detected, probably due to the inherent inhomogeneity in H distribution within the Pd samples.
While this protocol was tailored for Pd, selected here as a test material, it can potentially serve as a guideline for future IGP studies on various materials, including those of interest for hydrogen storage, high-temperature superconductivity\add{, and spintronics}.

\section*{Author contributions}
E.P.,  R.S.G. and D.D. conceived the idea.
R.S.G. directed the project.
E.P., R.S.G. and D.D. designed and performed the protonation and the electric transport measurements.
G.G.  performed the protonation, electric transport, and atomic force microscopy measurements and analyzed the data.
G.T., A.M., L.G. and C.S.P. designed the deep eutectic solvent.
G.T. and S.L. synthesized the deep eutectic solvent.
D.D.F. fabricated the palladium thin films.
G.G., E.P., R.S.G. and D.D. wrote the manuscript with input from all authors.


\section*{Acknowledgements}
\noindent
This work was supported by the MIUR PRIN-2017 program (Grant No.2017Z8TS5B – “Tuning and understanding Quantum phases in 2D materials – Quantum2D”). 
E.P., D.D.F. and R.S.G. acknowledge funding by the European Union - Next Generation EU as part of the PRIN 2022 PNRR project “Continuous THERmal monitoring with wearable mid-InfraRed sensors” (P2022AHXE5).
This publication is also part of the project PNRR-NGEU which has received funding from the MUR–DM 351/2022.
We thank J. Montagna Bozzone for assistance in the measurements.

\appendix
\section{Temperature correction for $\chi_\mathrm{R}(x)$}\label{appendix:flanagan_correction}
Since most of the representative studies depicted in Fig.\,\ref{fig:5}b map the dependence of $\chi_\mathrm{R}$ on $x$ at 295\,K, it is essential to extrapolate the data from Flanagan et al.\,\cite{Flangan_Lewis_1961}, originally obtained at 273\,K, to this common temperature to enable comparative analysis.
Assuming a linear relationship between $\rho$ and $T$ in this range, the expression for $\chi_\mathrm{R}$ as a function of $x$ at 295\,K can be derived as follows:
\begin{equation}
\chi_\mathrm{R} (x,T)= \frac{\rho(x,T_0)\left[1+\alpha_{PdH_x}(T-T_0)\right]}{\rho(0, T_0)\left[1+\alpha_{Pd}(T-T_0)\right]}
\label{eq:flanagan_rescaling}
\end{equation}
where $T=295$\,K, $T_0=\,273$\,K and $\alpha_{PdH_x}$ and $\alpha_{Pd}$ are the temperature coefficients of the resistivity for PdH$_x$ and Pd, respectively.
Solving this equation requires an analytical expression for $\alpha_{PdH_x}$ as a function of $x$.
Since the $\alpha_{PdH_x}-x$ data from Ref.\,\cite{Flangan_Lewis_1961} exhibit a linear trend, a linear interpolation of these data points is performed, resulting in the expression $\alpha_{PdH_x}(x) = (43 - 37x) \times 10^{-4}$\,K\apex{-1}.
Substituting it into Eq.\,\eqref{eq:flanagan_rescaling} and setting $\alpha_{Pd} = \alpha_{PdH_x}(0)$, we obtain the final relationship for extrapolating the $\chi_\mathrm{R}(x)$ data from Ref.\,\cite{Flangan_Lewis_1961} to 295\,K.\\

\bigskip

%

\end{document}